\documentclass[a4paper,fleqn,usenatbib,useAMS]{mnras}

\usepackage{graphicx}	
\usepackage{amsmath}	
\usepackage{amssymb}	
\usepackage{multicol}        
\usepackage{bm}		
\usepackage{pdflscape}	

\usepackage[dvipsnames]{xcolor}
\usepackage{tikz,hyperref}

\definecolor{lime}{HTML}{A6CE39}
\DeclareRobustCommand{\orcidicon}{
	\begin{tikzpicture}
	\draw[lime, fill=lime] (0,0) 
	circle [radius=0.13] 
	node[white] {{\fontfamily{qag}\selectfont \tiny ID}};
	\draw[white, fill=white] (-0.1625,0.00) 
	circle [radius=0.007];
	\end{tikzpicture}
	\hspace{-2.5mm}
}

\foreach \x in {A, ..., Z}{\expandafter\xdef\csname orcid\x\endcsname{\noexpand\href{https://orcid.org/\csname orcidauthor\x\endcsname}
			{\noexpand\orcidicon}}
}

\usepackage[T1]{fontenc}
\usepackage{ae,aecompl}

\usepackage{mathptmx}

\title[Morpho-kinematic study of NGC\,40]{Adjusting the bow-tie: A morpho-kinematic study of NGC\,40}

\author[Rodr\'{i}guez-Gonz\'{a}lez et al.]{J.~B. Rodr\'{i}guez-Gonz\'{a}lez\thanks{E-mail:\,j.rodriguez@irya.unam.mx}$^{1\orcidA}$, J.~A. Toal\'{a}$^{1\orcidB}$, L.~Sabin$^{2\orcidC}$, G.~Ramos-Larios$^{3\orcidD}$, M.~A.~Guerrero$^{4\orcidE}$ \newauthor{J.~A.\,L\'{o}pez$^{2\orcidF}$ and S.~Estrada-Dorado$^{1\orcidG}$}\\
$^{1}$Instituto de Radioastronom\'{i}a y Astro\'{i}sica, UNAM Campus Morelia, Apartado postal 3-72, 58090 Morelia, Michoac\'{a}n, Mexico\\
$^{2}$Instituto de Astronom\'{i}a, Universidad Nacional Aut\'{o}noma de M\'{e}xico (UNAM), Apdo. Postal 877, 22800 Ensenada, B.C., Mexico\\
$^{3}$Instituto de Astronom\'\i a y Meteorolog\'\i a, CUCEI, Univ.\ de Guadalajara, Av.\ Vallarta 2602, Arcos Vallarta, 44130 Guadalajara, Mexico\\
$^{4}$Instituto de Astrof\'{i}sica de Andaluc\'{i}a, IAA-CSIC, Glorieta de la Astronom\'{i}a S/N, Granada 18008, Spain
}

\pubyear{2021}

\begin{document}
\label{firstpage}
\pagerange{\pageref{firstpage}--\pageref{lastpage}}
\maketitle

\begin{abstract}
We present a comprehensive study of the ionization structure and kinematics of the planetary nebula (PN) NGC\,40 (a.k.a. the Bow-tie Nebula). A set of narrow-band images obtained with the ALhambra Faint Object Spectrograph and Camera (ALFOSC) at the Nordic Optical Telescope (NOT) are used to study the turbulent distribution of gas in the main cavity, the ionization stratification and the density of this PN. High-resolution Manchester Echelle Spectrograph (MES) observations obtained at 2.1m telescope of the San Pedro M\'{a}rtir (SPM) Observatory are used to unveil in great detail the kinematic signatures of all morphological features in NGC\,40. The images and spectra suggest that NGC\,40 had multiple mass ejections in its recent formation history. We found 4 jet-like ejections not aligned with the main axis of NGC\,40 (PA=20$^{\circ}$), some of them having pierced the main cavity along the SW-NE direction as well as the southern lobe. Using a tailor-made morpho-kinematic model of NGC\,40 produced with {\sc shape} we found that the main cavity has a kinematic age of 6,500~yr and the two pairs of lobes expanding towards the N and S directions have an averaged age of 4,100$\pm$550~yr. NGC\,40 thus adds to the group of PNe with multiple ejections along different axes that challenge the models of PN formation. 
\end{abstract}





\section{INTRODUCTION}
\label{sec:intro}

Planetary nebulae (PNe) are remnants of the late stages of stellar evolution of low- and intermediate-mass stars (1 $\lesssim$ $M_\mathrm{i}$/M$_\odot \lesssim$  8). These stars lose most of their initial masses when evolving through the asymptotic giant branch (AGB) phase where they exhibit a slow and dense wind \citep[$\Dot{M}\lesssim10^{-5}$ yr$^{-1}$; $v_{\mathrm{AGB}}\approx$ 20 km~s$^{-1}$;][]{Ramstedt2020}. They expel their outer layers and become post-AGB stars increasing their stellar temperature and developing thus a strong line-driven wind \citep[$v_\infty$ = 500--4000 km~s$^{-1}$;][]{Guerrero2013} that compresses and heats the material previously ejected in the AGB stage. At the same time, they develop a UV photon flux that ionizes the material, giving birth to a PN \citep{Kwok1978,Balick1987}.

About $\sim$20 per cent of the known PNe harbor central stars (CSPN) that do not display any hints of H in their atmospheres \citep{Weidmann2020}. These H-deficient stars exhibit strong emission lines of He, C and O, very similar to the massive Wolf-Rayet (WR) stars of the C sequence \citep[WC; e.g.,][]{Acker2003}. Therefore they are also classified under the same scheme, but using square brackets to distinguish them from their massive siblings \citep{Crowther1998}.

PNe harboring WR-type CSPN (hereinafter WRPNe) display some differences when compared to PNe harboring H-rich CSPN. Generally, they seem to have higher N and C abundances suggesting that they are the descendants of more massive stars \citep[see][]{GarciaRojas2013}, which means that they are preferentially found close to the Galactic plane \citep{Corradi1995,Pena2013}. Some authors have found that WRPNe exhibit a larger degree of turbulence and present larger expansion patterns than non-WRPNe \citep[e.g.,][]{Medina2006,Jacob2013}. These properties are easily explained in WRPNe by invoking the large momentum imprinted by the powerful line-driven winds from WR-type CSPN \citep{Todt2015}. They compress and accelerate the previously ejected AGB material much stronger than those of H-rich CSPN, enhancing the formation of hydrodynamical instabilities such as the Rayleigh-Taylor \citep[e.g.,][]{Stute2006,Toala2016}.

\begin{figure*}
\begin{center}
\includegraphics[width=0.95\linewidth]{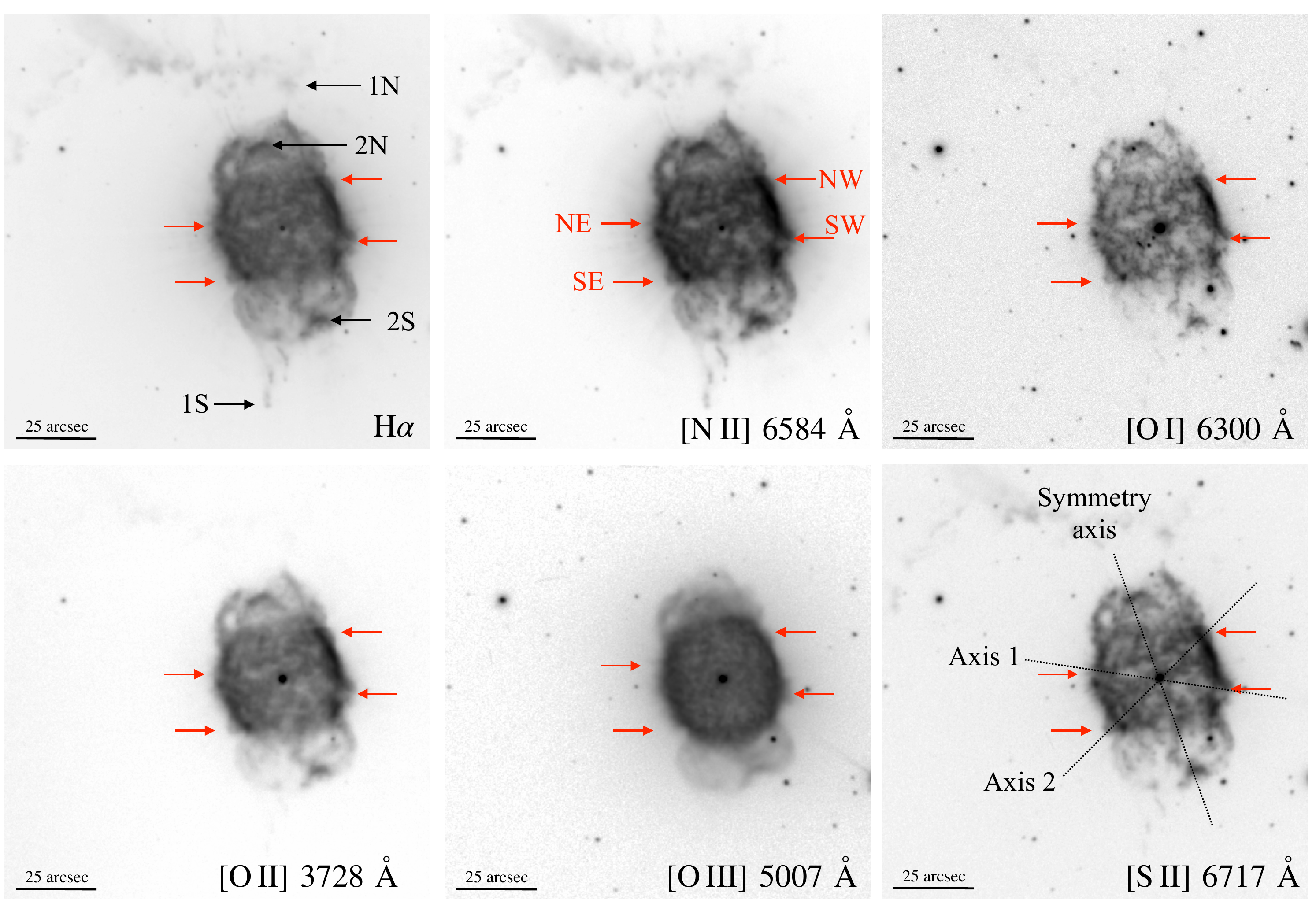}
\caption{Narrow-band images of NGC\,40 obtained with NOT ALFOSC. Each panel presents images in different emission lines. The position of the possible jet-like features 1N, 2N, 1S, and 2S are marked in the upper left panel with black arrows. The two pairs of blowouts piercing the main cavity of NGC\,40 are shown with red arrows in all panels (see text for details). The bottom-right panel shows the symmetry axis of NGC\,40 (PA=20$^{\circ}$). Axis 1 (PA=80$^{\circ}$) and 2 (PA=135$^{\circ}$) are defined by connecting the two pairs of blowouts protruding from the main cavity of NGC\,40. All panels have the same field of view. North is up, east to the left.} 
\label{fig:figt}
\end{center}
\end{figure*}

Kinematic studies of PNe help us assess the effects of the powerful winds from their CSPNe, investigate the mass-loss ejection histories that gave birth to these objects and their interaction with the circumstellar medium \citep[see, e.g.,][and references therein]{Aller2021,Sabin2017,RamosLarios2018,Henney2021,Guerrero2021}. In this paper we present a morpho-kinematic study of the WRPN NGC\,40, also known as the Bow-tie Nebula, around the [WC]-type star HD\,826 \citep[][]{Hiltner1966}.

The main structure of NGC\,40 has a barrel-like shape with two pairs of lobes expanding towards the northern and southern directions (see Fig.~\ref{fig:figt}). Its symmetry axis appears to have a position angle (PA) of $20^{\circ}$ \citep[][]{Chu1987,Sabbadin1982}. Ring-like features detected in optical and IR images have been reported to surround the main structure of NGC\,40 \citep{Corradi2004,RamosLarios2011}. Deep optical images of NGC\,40 reveal an extended clumpy halo that extends more than 2.2~arcmin  \citep[see fig.~1 in][]{Toala2019}. Kinematical studies of NGC\,40 have suggested that its bipolar appearance is due to the possible action of bipolar ejections of material, jet-like features, oriented at PA$\approx0^{\circ}$ and PA=20$^{\circ}$ \citep{Sabbadin1982,Meaburn1996}.

In this paper we present a 3D morpho-kinematic study of NGC\,40 with the goal of unveiling the kinematical signature of possible different ejections from HD\,826. High-resolution spectroscopic observations are used to peer into the formation scenario of NGC\,40. These are interpreted by means of the 3D modelling tool for Astrophysics {\sc shape} \citep{Steffen2011}. We also investigate the ionization structure of NGC\,40 by means of narrow-band imagery. This paper is organized as follows. In Section~2 we present the observations used in the present paper. In Section~3 we describe our results. The morpho-kinematic {\sc shape} model is presented and compared to the observations in Section~4. The discussion and conclusions are presented in Sections~5 and 6, respectively.

\section{Observations}

We obtained optical images and spectra of NGC\,40 from different telescopes.
In the following we describe the details of the observations. We note that all observations presented in this paper were processed using standard {\sc iraf} routines \citep{Tody1993}.

\subsection{NOT ALFOSC images}

\begin{figure*}
\begin{center}
\includegraphics[angle=0,width=0.9\linewidth]{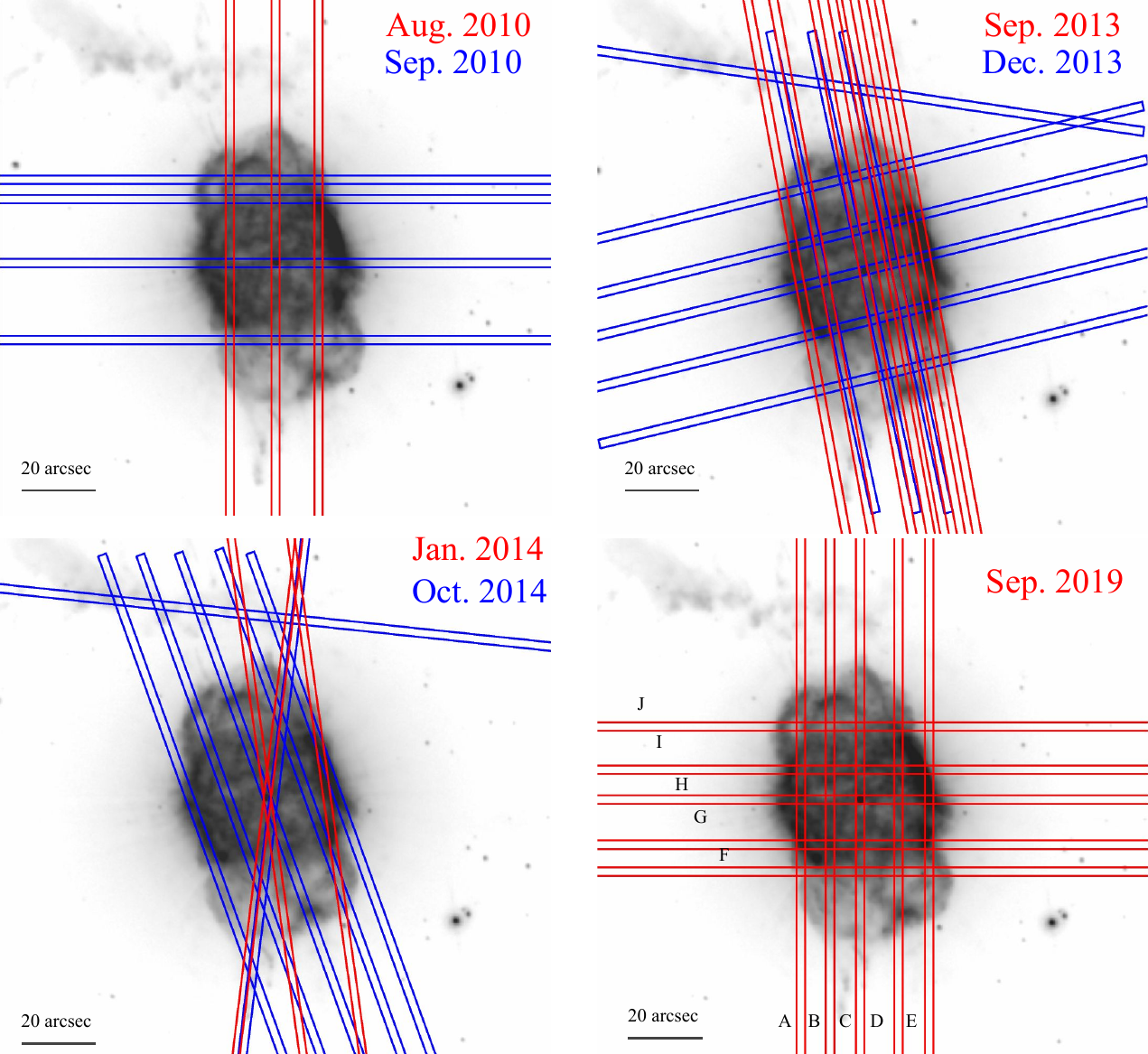}
\caption{Narrow-band [N\,{\sc ii}] image of NGC\,40 with the slit positions of the MES observations overplotted. The upper and lower-left panels show all the slits obtained with the spectroscopic configuration covering the H$\alpha$ and [N\,{\sc ii}] emission lines with different colours illustrating different observing runs. Similarly, the lower right panel shows examples of the slit positions along key nebular features with the spectroscopic configuration covering the [S\,{\sc ii}] doublet.}
\label{fig:MES}
\end{center}
\end{figure*}

\begin{table}
\caption{Details of the narrow-band filters used for the NOT ALFOSC observations.}
\centering
\begin{tabular}{lccc} 
\hline
Filter   & $\lambda_\mathrm{c}$ & FWHM   & Exposure time \\
         & (\AA)                & (\AA)  & (s) \\
\hline
[O\,{\sc ii}]   &  3728  & 32    &  600 \\ 

H$\beta$        &  4854  &  133  & 1200 \\

[O\,{\sc iii}]  &  5007  & 30    & 1200 \\

[O\,{\sc i}]    &  6300  & 30    & 1200 \\

H$\alpha$       &  6564  & 3.3   & 600 \\

[N\,{\sc ii}]   &  6584  & 10    & 450 \\

[S\,{\sc ii}]   &  6716  & 10    & 1200 \\

[S\,{\sc ii}]   &  6731  & 10    & 1200 \\
\hline
$g'$            &  4800  & 1450 & 30 \\
$r'$            &  6250  & 1400 & 30 \\
\hline
\end{tabular}
\label{tab:alfosc}
\end{table}

NGC\,40 was observed on 2019 October 11 (Proposal ID 60-208, PI: M.A.\,Guerrero) with the Alhambra Faint Object Spectrograph and
Camera (ALFOSC)\footnote{\url{http://www.not.iac.es/instruments/alfosc/}} 
mounted at the Nordic Optical Telescope (NOT) at the Observatorio de El Roque de los Muchachos (ORM) in La Palma (Spain). Images were obtained using narrow-band filters centred on the H$\alpha$, H$\beta$, [O\,{\sc i}]$\lambda$6300, [O\,{\sc ii}]$\lambda$3727, [O\,{\sc iii}]$\lambda$5007, [N\,{\sc ii}]$\lambda$6584 and [S\,{\sc ii}]$\lambda\lambda$6716,6731 emission lines. Broad-band images on the $r'$ and $g'$ filters were also obtained. Details of the observations such as filter, central wavelength ($\lambda_\mathrm{c}$), FWHM and total exposure time per filter are listed in Table~\ref{tab:alfosc}. A selection of narrow-band images is presented in Fig.~\ref{fig:figt}.

\subsection{SPM MES spectra}

NGC\,40 was observed with the Manchester Echelle Spectrograph \citep[MES;][]{Meaburn2003} at the 2.1m telescope of the Observatorio
Astron\'{o}mico Nacional (OAN) in San Pedro M\'{a}rtir (SPM) 
in Ensenada, México. The observations were performed through several runs
between 2010 and 2019. A total of 60 high-resolution MES spectra were obtained at different
positions and with different filters. 33 spectra were acquired with the H$\alpha +$[N\,{\sc ii}] interference filter with $\lambda_\mathrm{c}$=6580 and $\Delta \lambda$=90~\AA. This filter isolates the 87th order which includes the H$\alpha$, He\,{\sc ii}~6560~\AA, [C\,{\sc ii}]~6578~\AA\ and the [N\,{\sc ii}]~6548,6584~\AA\ doublet emission lines. In addition we also obtained 27 spectra through the [S\,{\sc ii}] filter with $\lambda_\mathrm{c}$=6730~\AA\, and $\Delta \lambda$=90~\AA\, that includes the [S\,{\sc ii}]6716,6731~\AA\, doublet. Details of the observations are listed in Table~\ref{tab:MES}. The slit positions are illustrated in Fig.~\ref{fig:MES}.

\begin{table}
\caption{Details of the spectroscopic data.}
\centering
\begin{tabular}{llccccc} 
\hline
Instrument & Filter    & $\lambda$ & $\Delta \lambda$ & Season   & \# of  & Exp. Time \\
           &           & (\AA)     & (\AA)        &  & Spectra & (s)       \\
\hline
MES SPM & H$\alpha$    & 6580 & 90 & Aug 2010 & 3           & 1200          \\
        &              & 6580 & 90 & Sep 2020 & 4           & 1200          \\
        &              & 6580 & 90 & Sep 2013 & 7           & 1800          \\
        &              & 6580 & 90 & Dec 2013 & 9           & 1800 \\
        &              & 6580 & 90 & Jan 2014 & 3           & 1800 \\
        &              & 6580 & 90 & Oct 2014 & 7           & 1800 \\
        &[S\,{\sc ii}] & 6730 & 90 & Sep 2019 & 9           & 1800  \\
        &              & 6730 & 90 & Oct 2019 & 18          & 1800  \\

\hline
WHT UES & H$\alpha$ & 6590 & 130  & Oct 1995  & 1 & 600 \\
        & [O\,{\sc i}] & 6303 & 50 & Oct 1995  & 1 & 600 \\
        &  [O\,{\sc ii}] & 3737 & 50 & Oct 1995  & 1 & 600 \\
        & [O\,{\sc iii}] & 5010 & 100 & Oct 1995  & 1 & 600 \\
\hline
\end{tabular}
\label{tab:MES}
\end{table}

\subsection{WHT UES spectra} 

To complement our SPM-MES spectra, we also retrieved archival Utrech Echelle 
Spectrograph (UES) observations\footnote{The observations were retrieved from the Isaac Newton Group Archive at \url{http://casu.ast.cam.ac.uk/casuadc/ingarch/query}}. The instrument was mounted on the William Herschel Telescope (WHT) also at ORM. The observations were obtained using a Tek~5 CCD detector with a spectral resolution of $R$=49,000 which corresponds to a velocity resolution of $\sim$6~km~s$^{-1}$.

The WHT UES observations were acquired in 1995 October 11 using the 3720, 5005, 6300, 6570 and 6705~\AA\ filters, which include the H$\alpha +$[N\,{\sc ii}], [O\,{\sc i}], [O\,{\sc iii}] emission lines  and the [O\,{\sc ii}]~3726,3728~\AA\, and [S\,{\sc ii}]~6716,6731~\AA\, doublets. All observation were taken with the same slit position covering the symmetry axis of NGC\,40 (PA=20$^{\circ}$) across the CSPN. Observations details are also presented in Table~\ref{tab:MES}.

\section{RESULTS}

\subsection{Imaging}

The narrow-band images of NGC\,40 presented in Fig.~\ref{fig:figt} exhibit all the morphological features described in previous works \citep[see fig.~3 in][]{Meaburn1996}: the main barrel-like cavity, the blister-like structures
expanding towards the N-S with PA=20$^{\circ}$, and the northern filament extending along E-W, this one detected in the H$\alpha$, H$\beta$ (not shown here), [N\,{\sc ii}] and [S\,{\sc ii}]. Some marginal contribution to the [O\,{\sc i}] narrow-band image is also detected.

Almost all narrow-band images have very similar morphologies, except for the [O\,{\sc i}] and [O\,{\sc iii}], which trace the lowest and highest ionization structures in NGC\,40, respectively. As discussed in \citet{Meaburn1996}, the [O\,{\sc iii}] image has a more organized clumpy morphology than show in other filters, which are messier or more complex. Furthermore, the internal cavity detected in this image has a smaller extent \citep[see fig.~11 in][]{Meaburn1996}. The [O\,{\sc iii}] image can be fitted by an ellipse with a semi-minor axis of 18~arcsec, whilst for the other narrow-band images an ellipse with a semi-minor axis of 20~arcsec is necessary.

\begin{figure*}
\includegraphics[width=\textwidth]{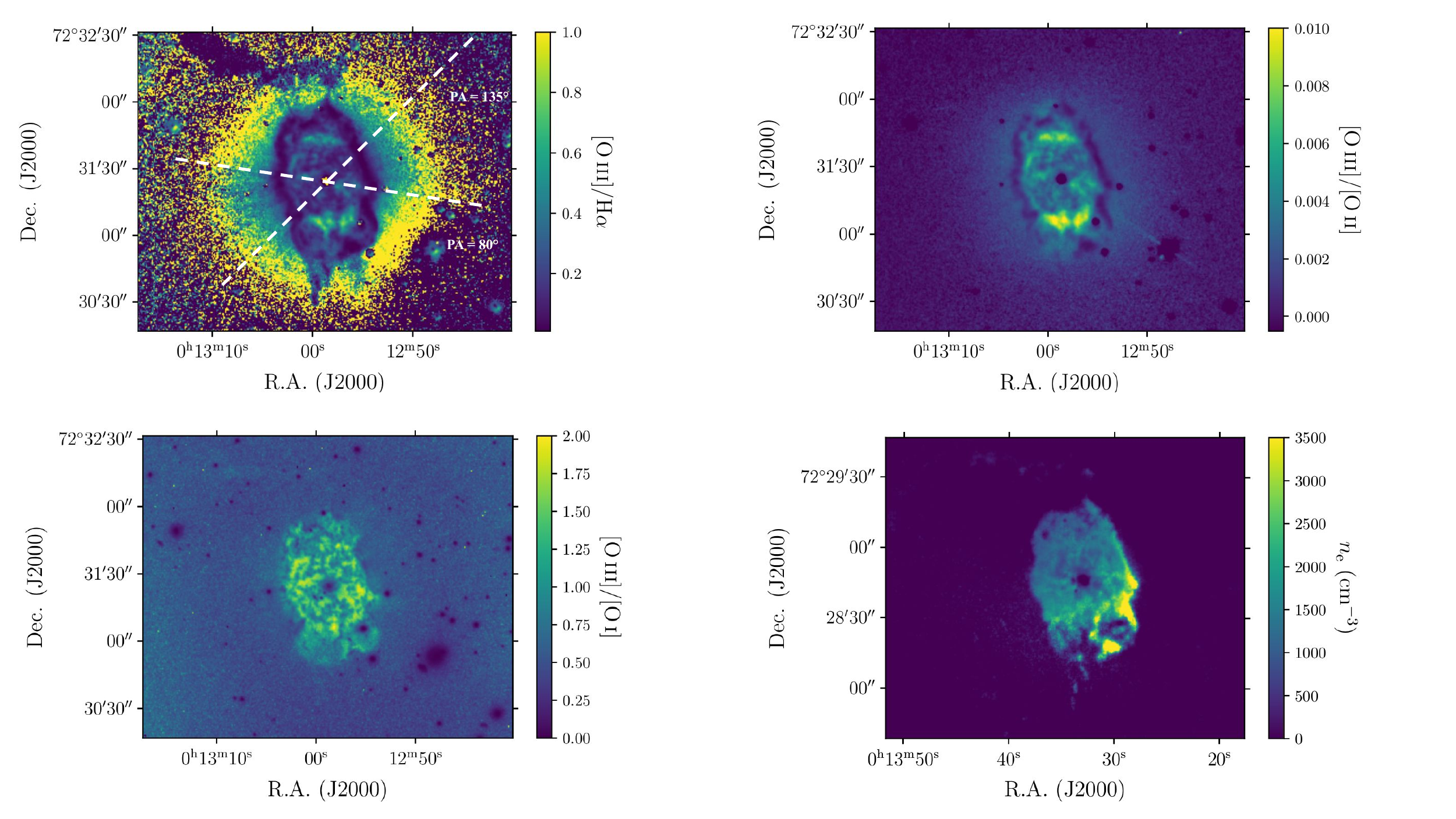}
\caption{Ratio maps of different continuum-subtracted narrow-band images of NGC\,40. The panels show the [O\,{\sc iii}]/H$\alpha$ (top-left), [O\,{\sc iii}]/[O\,{\sc ii}] (top-right), [O\,{\sc iii}]/[O\,{\sc i}] (bottom-left). The bottom-right panel presents the electron density ($n_\mathrm{e}$) map obtained from the [S\,{\sc ii}]~6716,6731~\AA\, images using {\sc PyNeb}.}
\label{fig:ratios}
\end{figure*}

The subtle morphological differences presented by the different narrow-band images of NGC\,40 can be explained as the result of the combination of the current fast stellar wind \citep[$v_{\infty}$=1000~km~s$^{-1}$;][]{Toala2019} and the ionizing UV flux. The current fast wind is sweeping the inner nebular material producing Rayleigh-Taylor instabilities \citep[see, e.g.,][]{Stute2006,Toala2016}, creating a more or less organized inner clumpy structure and at the same time the UV flux from HD\,826 ionizes the material. The ionization flux is trapped in the densest knots producing the so-called shadowing instability \citep{Williams1999,Arthur2006}. As a result, the ionization is not uniform through NGC\,40 creating a more chaotic pattern of ionized structures for larger radii such as the low-ionization structure detected in the [O\,{\sc i}] image. This is consistent with the streaks of alternating ionized material observed in the halo just outside the main cavity of NGC\,40 revealed in deep optical and IR images \citep[][]{Corradi2004,RamosLarios2011}.

An inspection of the narrow-band images disclose the presence of two pairs of blowouts located on the E and W walls of the main cavity of NGC\,40. Their positions are marked in all panels of Fig.~\ref{fig:figt} with red arrows. The most evident are those blowouts located at the NW, SW, and SE edges of the central cavity. However, a fourth blowout can be identified after careful examination of the images on the NE part of this region. The pairs of blowouts break the main barrel like structure of NGC\,40 at PA$\approx$80$^{\circ}$ and 135$^{\circ}$. This situation is illustrated in the bottom-right panel of Fig.~\ref{fig:figt}. It is interesting to note that the pair of blowouts in Axis 1 (PA=80$^{\circ}$) seem to have pierced through NGC\,40 while those in Axis~2 (PA=135$^{\circ}$) seem to be currently disrupting the edge of the main cavity. Interestingly, the maxima of the X-ray emission detected by {\it Chandra} are also aligned with Axis~2 \citep[see][]{Montez2005}\footnote{See also \url{https://chandra.harvard.edu/photo/2005/n40/}}.

To study the relative variation of excitation and to assess the presence of shocked structures in NGC\,40, we have created line ratio images from different narrow-band images. Before producing the ratio maps, we subtracted the continuum using the $g'$ and $r'$ broad band images of NGC\,40. Hence, for images obtained at longer wavelengths (H$\alpha$, [N\,{\sc ii}], [O\,{\sc i}] and [S\,{\sc ii}]) we have subtracted the $r'$ filter, whilst for the rest of the images (H$\beta$, [O\,{\sc ii}] and [O\,{\sc iii}]) we have subtracted the $g'$ broad filter. Examples of the ratio maps are presented in Fig.~\ref{fig:ratios}. We note that although the images are not flux-callibrated, they can still be used to describe in detail the ionization structure of NGC\,40.

A fast outflow propagating into a low-density medium generates a forward shock that produces an increase in the electron temperature.  
This would increase the emission in the [O\,{\sc iii}]~5007~\AA\ line, which is sensitive to $T_\mathrm{e}$ changes, while the low density would reduce the emissivity of the H$\alpha$ line, which is more sensitive to $n_\mathrm{e}$. 
Consequently, the ratio [O\,{\sc iii}]/H$\alpha$ is expected to increase significantly in regions of shocked gas \citep{Guerrero2008, Guerrero2013b}.
The [O\,{\sc iii}]/H$\alpha$ ratio map (Fig.~\ref{fig:ratios} top-left panel) shows that NGC\,40 is delimited by a structure with relatively low values. Almost the entire nebula can be delineated with values [O\,{\sc iii}]/H$\alpha < 0.2$, which is due to the ionization stratification in NGC\,40 with the [O\,{\sc iii}] structure contained within the H$\alpha$-emitting region. 
There are certain directions, however, where the values of the [O\,{\sc iii}]/H$\alpha$ ratio increase, particularly along the direction of the SW blowout. 

Within the main cavity of NGC\,40 there are areas with values [O\,{\sc iii}]/H$\alpha >$0.6, two corresponding to the N and S areas as well as an elongated structure that extends from the NW to the SE of the inner cavity where the blowouts align with Axis~1 (PA=80$^{\circ}$). A similar ionization structure is unveiled by the [O\,{\sc iii}]/[O\,{\sc ii}] map (Fig.~\ref{fig:ratios} top-right panel).

An unprecedented view of NGC\,40 is presented by the [O\,{\sc iii}]/[O\,{\sc i}] ratio map (Fig.~\ref{fig:ratios} bottom-left panel), which unveils its inner clumpy morphology. This image also discloses material extending beyond the main cavity of NGC\,40, in particular the SW blowout along PA 260$^\circ$, which traces ionized material streaking out from the western edge of the main cavity. The latter effect can not be appreciated in the western region of the main cavity.

\begin{figure*}
\begin{center}
\includegraphics[angle=0,width=0.95\linewidth]{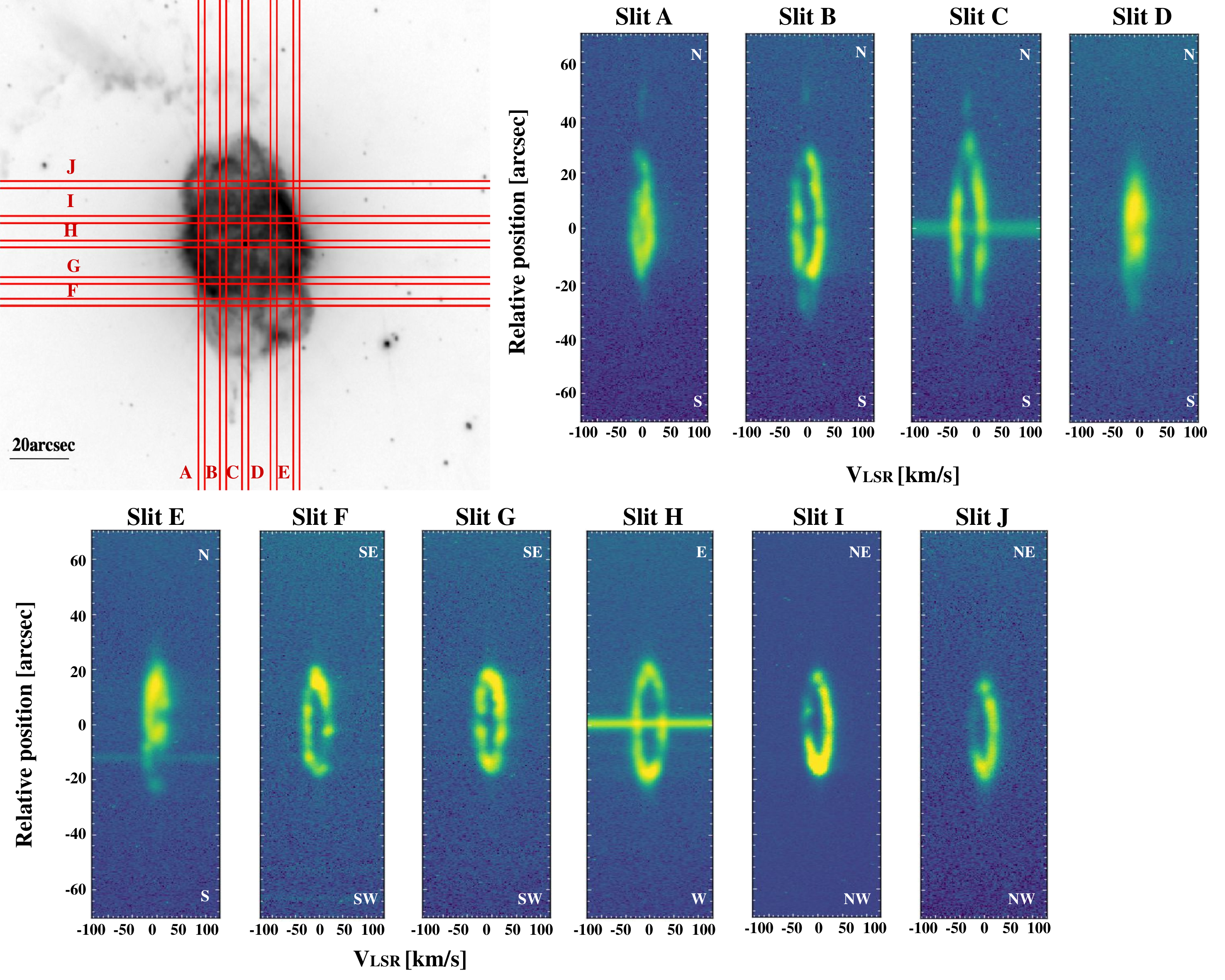}
\caption{SPM MES [N\,{\sc ii}]~6584~\AA\, spectra of NGC\,40 obtained during December 2013.}
\label{fig:MES_Dic2013}
\end{center}
\end{figure*}

\begin{figure*}
\begin{center}
\includegraphics[angle=0,width=0.95\linewidth]{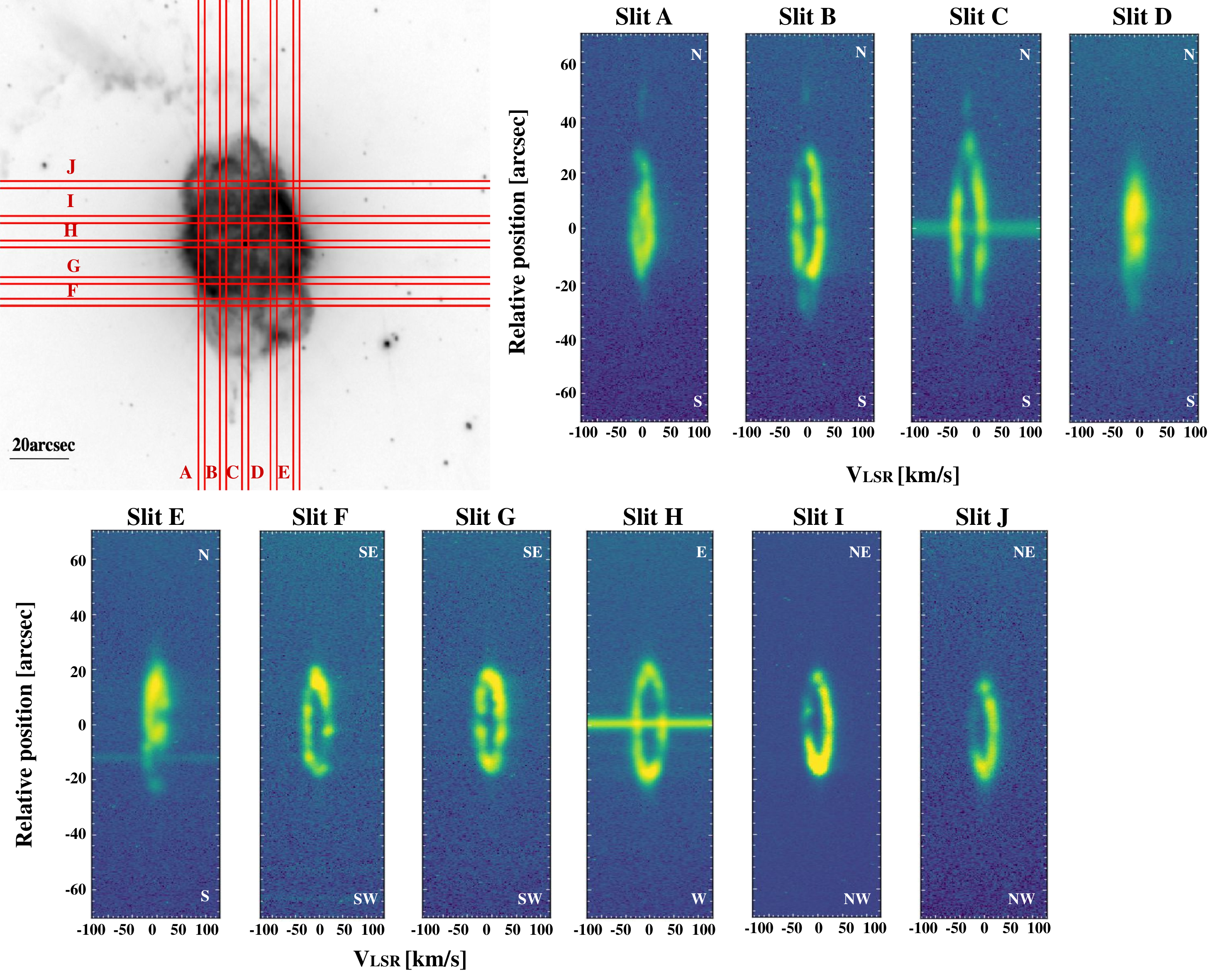}
\caption{SPM MES [S\,{\sc ii}]~6716~\AA\, spectra of NGC\,40 obtained during September and October 2019.}
\label{fig:SII_PV}
\end{center}
\end{figure*}

Finally, since we also have continuum-subtracted images of the [S\,{\sc ii}]~6716,6731~\AA\, doublet, we calculated an electron density ($n_\mathrm{e}$) map. For this, we used the {\sc PyNeb} task {\it getTemDen} \citep[][]{Luridiana2015} adopting the electron temperature of $T_\mathrm{e}=$8,200~K as reported by \citet{Pottasch2003}. Fig.~\ref{fig:ratios} bottom-right panel presents the resultant $n_\mathrm{e}$ map which shows that NGC\,40 has values around 1000 and 2000~cm$^{-3}$ with some areas with higher $n_\mathrm{e}$ found in the SW regions and indicating electronic densities $n_\mathrm{e}\gtrsim$3,500 cm$^{-3}$. Our $n_\mathrm{e}$ map is consistent with the $n_\mathrm{e}$ values reported in previous works \citep[see, e.g.,][and references therein]{Clegg1983,Pottasch2003,LealF2011}.

\subsection{Spectra}

\begin{figure*}
\begin{center}
\includegraphics[width=\linewidth]{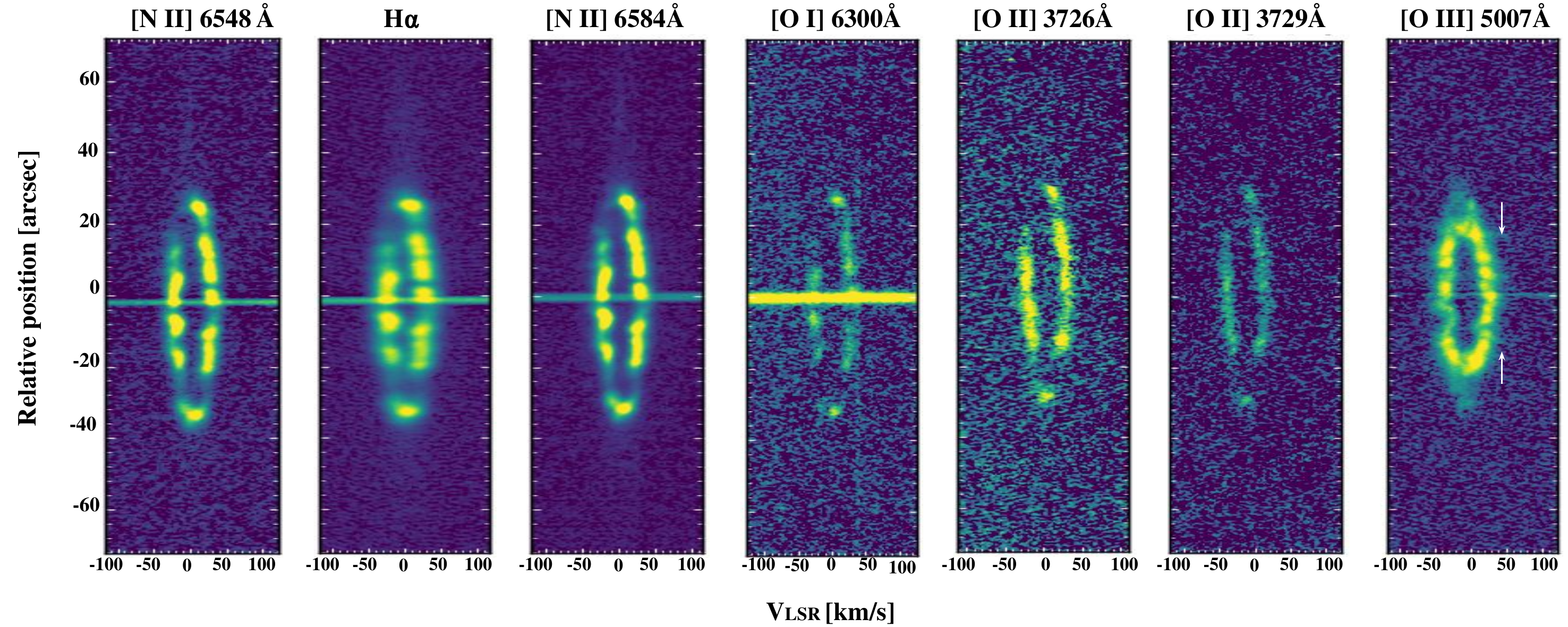}
\caption{WHT UES spectra of NGC\,40 taken with different filters. 
All spectra were obtained with a PA=20$^{\circ}$. Arrows in the [O\,{\sc iii}] PV (rightmost panel) indicate high velocity components (see Fig.~\ref{fig:OIII_UES} and the discussion section for details).}
\label{fig:UES}
\end{center}
\end{figure*}

\begin{figure}
\begin{center}
\includegraphics[width=0.6\linewidth]{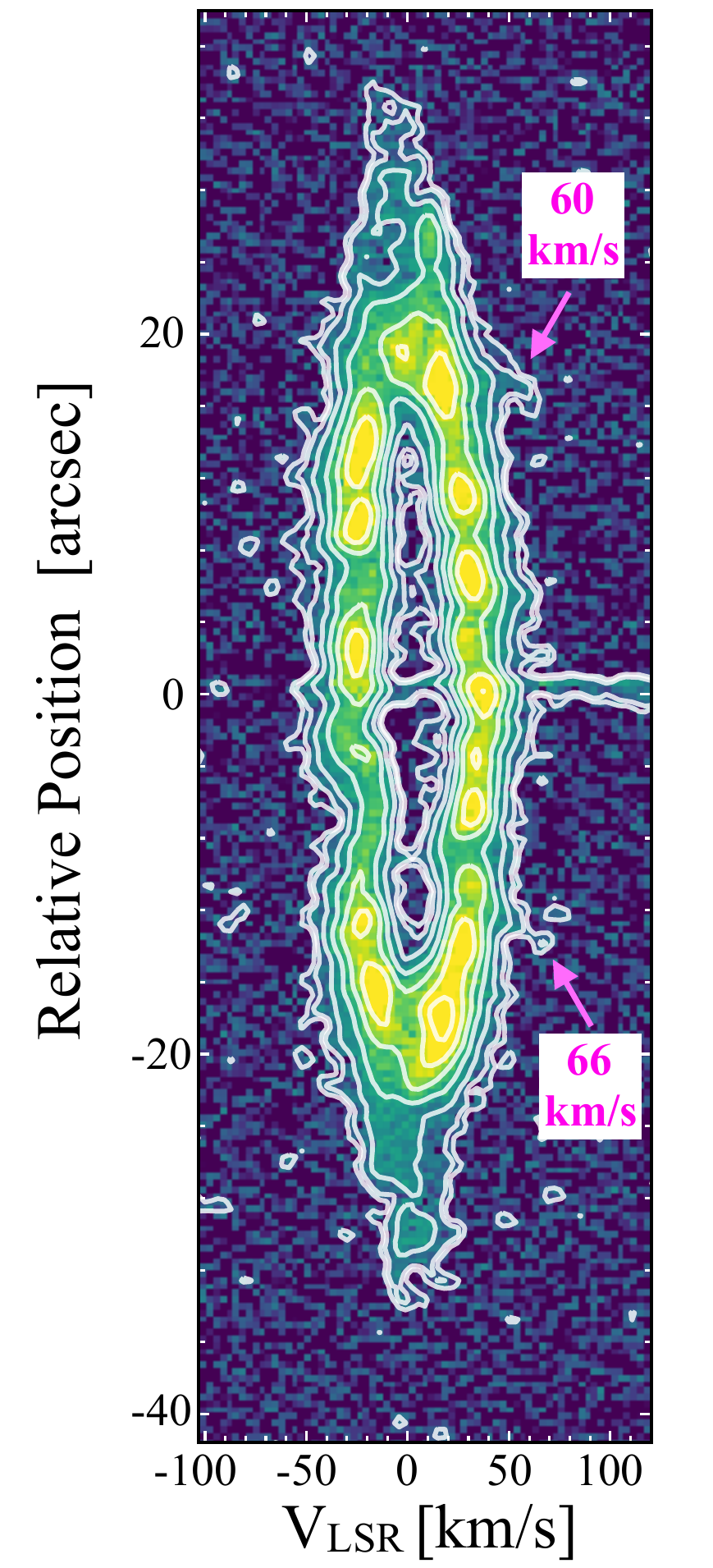}
\caption{WHT UES [O\,{\sc iii}] 5007~\AA\ PV of NGC\,40 at PA=20$^{\circ}$. The (magenta) arrows show the high-velocity structures. The northern and southern structures have velocities of 60 and 66~km~s$^{-1}$, respectively.}
\label{fig:OIII_UES}
\end{center}
\end{figure}

We created Position-Velocity (PV) diagrams for all spectra presented in this paper, although only a selection of the large number of PV diagrams derived from the SPM-MES observations are shown for simplicity. In Fig.~\ref{fig:MES_Dic2013} we present the [N\,{\sc ii}]~6584~\AA\ PV diagrams corresponding to observations performed on December 2013, while in Fig.~\ref{fig:SII_PV} we present those of the [S\,{\sc ii}]~6716~\AA\, obtained on September and October 2019. Fig.~\ref{fig:UES} shows the corresponding PV diagrams from the WHT UES observations. All spectra were corrected from Local Standard Rest (LSR) using the {\sc iraf} {\it rvcorrect} task and are plotted with respect to systemic velocity, which was determined to be $-$25 km~s$^{-1}$ using H$\alpha$ and [N\,{\sc ii}] spectra profile at the position of the central star.

In order to obtain expansion velocities ($v_{\mathrm{exp}}$) we performed Gaussian fits to the different components in the PV diagrams. The expansion velocity obtained from the [N\,{\sc ii}] spectra taken along the minor axis of the central cavity of NGC\,40, for example slit C in Fig.~\ref{fig:MES_Dic2013}, resulted in $v_{\mathrm{exp}}$=25~km~s$^{-1}$. Spectra obtained using the symmetry axis, such as slit G in Fig.~\ref{fig:MES_Dic2013}, resulted in an expansion velocity $v_{\mathrm{exp}}$=22~km~s$^{-1}$. The expansion velocity of the northern lobes were found to be $v_{\mathrm{exp}}$=13~km~s$^{-1}$, while the ones in the southern region of the main cavity resulted in $v_{\mathrm{exp}}$=10~km~s$^{-1}$. The PV of the filamentary structure located at the north from NGC\,40 (slit I in Fig.~\ref{fig:MES_Dic2013}) does not show a relevant velocity. Most of this filament appears to be located at the same velocity of the CSPN of NGC\,40 ($v_\mathrm{LSR}\approx$0~km~s$^{-1}$) with a small velocity gradient of 5~km~s$^{-1}$ from the NW part  of this structure.

The UES spectra were obtained for different emission lines at the same PA = 20$^\circ$ (the symmetry axis), therefore, giving us the advantage of studying the kinematics of NGC\,40 taking into account the ionization structure. Spectra of H$\alpha$, [N\,{\sc ii}]$\lambda\lambda$6548,6584, [O\,{\sc i}]$\lambda\lambda$6300,6363, [O\,{\sc ii}]$\lambda$3726,3729, and [O\,{\sc iii}]$\lambda\lambda$4957,5007 emission lines were obtained. The profiles have similar shapes except for the [O\,{\sc iii}] emission lines, similarly to the situation with the narrow-band images, the later exhibits a clumpier structure and it is definitely contain within the rest of the structures. An expansion velocity of this structure of 30~km~s$^{-1}$ was obtained for the spectrum of [O\,{\sc iii}], while expansion velocities of 24 and 26~km~s$^{-1}$ were obtained for the [N\,{\sc ii}] and [O\,{\sc ii}] spectra, respectively. The slowest expansion is seen in the spectrum of [O\,{\sc i}] with 17.7~km~s$^{-1}$, which corresponds to the outermost structures in NGC\,40.

It is interesting to mention that we detect a pair of red-shifted high-speed components in the [O\,{\sc iii}] PV diagram projected onto the main cavity.
We marked these components with (white) arrows in the rightmost panel of Fig.~\ref{fig:UES}. 
A close up of the [O\,{\sc iii}] PV diagram presented in Fig.~\ref{fig:OIII_UES} shows that they have velocities of 60~km~s$^{-1}$ for the northern component and 66~km~s$^{-1}$ for the southern one, located 17~arcsec and 13.8~arcsec from the central star, respectively. 
Additional high-velocity features might be present at other spatial locations in the [O\,{\sc iii}] PV diagram, but their low signal-to-noise (S/N) renders their detection questionable.

\section{A morpho-kinematic model}
\label{sec:shape}

To interpret the SPM MES observations in order to determine the true structure of NGC\,40 and to investigate its formation history, we have used {\sc shape}. This software is used to reconstruct and model the morpho-kinematic signatures of astrophysical objects. This procedure will be applied to NGC\,40 by comparing synthetic PV diagrams and images with those obtained from observations. 
We note that the bulk of material is detected through the H$\alpha$ and [N\,{\sc ii}] emission lines, but the latter suffers from much less thermal broadening. Thus, we will use {\sc shape} to create synthetic PV diagrams that will be directly compared with those obtained from the [N\,{\sc ii}] $\lambda$6584 \AA\ SPM MES spectra, which is then assumed to describe the general formation scenario of NGC\,40.

We started by including a barrel-like component to reproduce the main structure. The PA of the main axis of the barrel-like structure is chosen to be 20$^{\circ}$, as suggested by the symmetry axis of NGC\,40. This structure has a semi-major and semi-minor axis of 13 and 20~arcsec with a thickness of 8~arcsec. To reproduce the PV diagrams crossing over the main cavity of NGC\,40, the {\sc shape} model requires the barrel-like structure to have an expansion velocity of 25~km~s$^{-1}$.

Given the complexity of NGC\,40 revealed by the narrow-band images, we added additional structures. The northern and southern lobes protruding from the main structure are modelled by including two pairs of cap structures.
The northern and southern pairs of caps are modelled using truncated spheres. The northern pair of caps have averaged radii of 8 with thickness of 7~arcsec. The southern pair of caps are slightly larger than the northern ones, with averaged radii of 10~arcsec and thickness of 5~arcsec. To reproduce the SPM MES spectra passing through the lobes, the model required expansion velocities of the northern and southern caps of 26 and 15~km~s$^{-1}$, respectively. Details of these structures are listed in Table~\ref{tab:ages}.

\begin{figure*}
\begin{center}
\includegraphics[width=\linewidth]{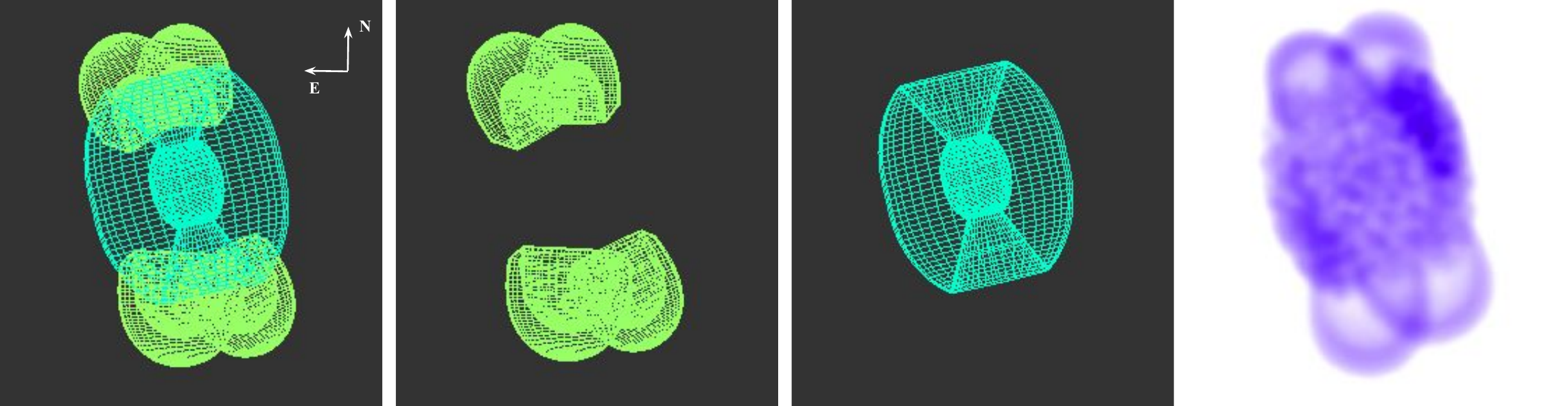}
\caption{Components used in {\sc shape} model to reproduce the SPM-MES spectra of NGC\,40. The rightmost panel shows a rendered synthetic image of the model.}
\label{fig:shape}
\end{center}
\end{figure*}

We present in Fig.~\ref{fig:shape} the visualization of our {\sc shape} model. The clumpy structure of NGC\,40 unveiled by the narrow-band images are simulated adding noise to the model. Additional clumps have been added in the model to the NW and SE regions of the barrel-like main cavity to reproduce its higher brightness. The synthetic image of NGC\,40 created from the morpho-kinematics reconstruction makes a good job reproducing most of the narrow-band images presented in previous sections.

\begin{figure*}
\begin{center}
\includegraphics[width=\linewidth]{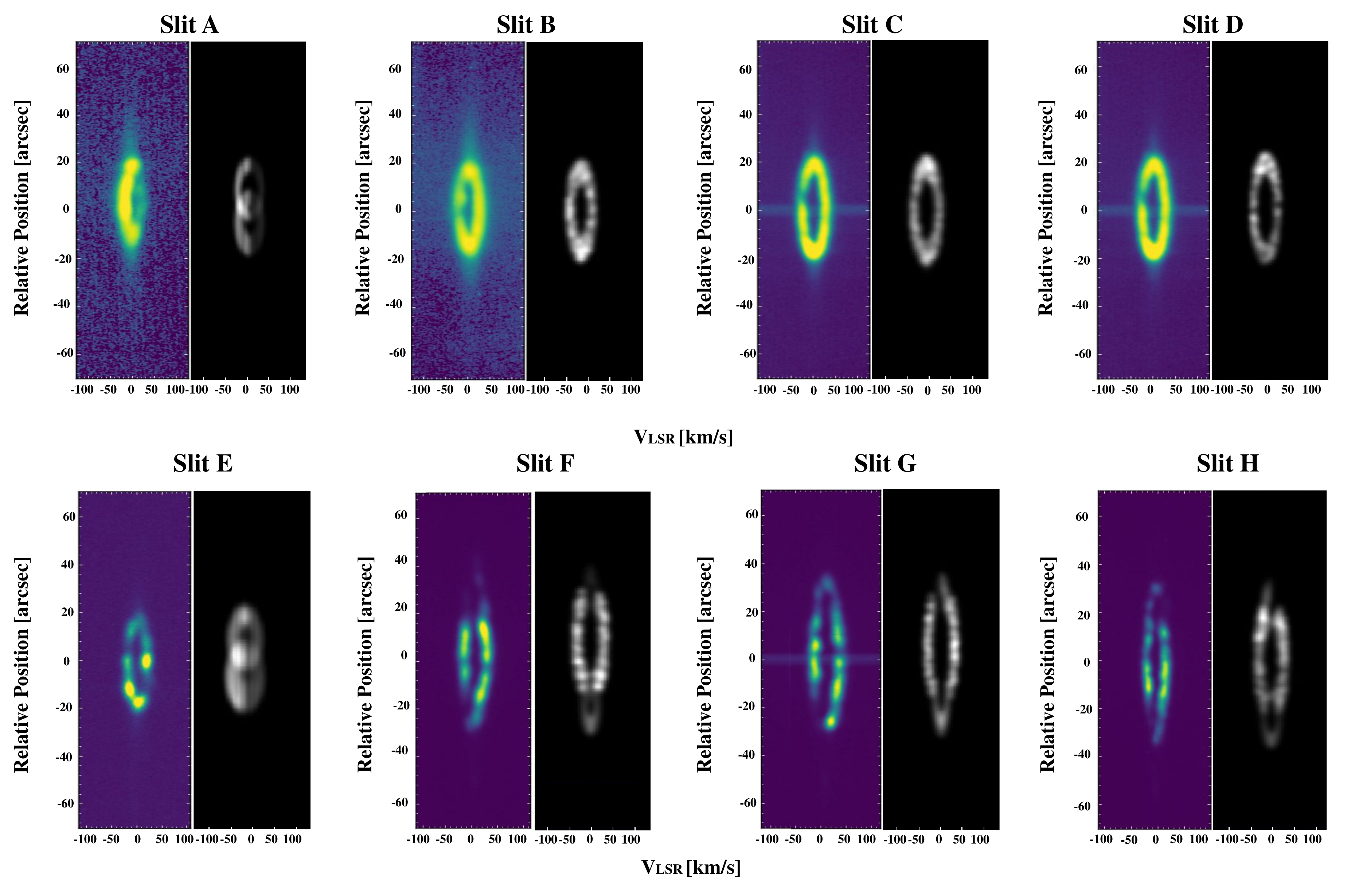}
\caption{Synthetic PV diagrams obtained from our {\sc shape} model of NGC\,40 (grey-scaled panels) compared with SPM-MES [N\,{\sc ii}] 6584~\AA\ observations obtained on December 2013 and shown in Figure~\ref{fig:MES_Dic2013} (colored panels).}
\label{fig:shape_sinteticos}
\end{center}
\end{figure*}

Synthetic PV diagrams obtained from our {\sc shape} model of NGC\,40 are presented in Fig.~\ref{fig:shape_sinteticos}. 
These are compared side-by-side with the SPM MES observations obtained on December 2013, which are representative of the kinematics of NGC\,40 as they cover its main structures. Our model reproduces fairly well the observed [N\,{\sc ii}] $\lambda$6584 \AA\ PVs obtained from different slit positions.

\begin{table}
\caption{Details of the different components from NGC\,40.}
\centering
\begin{tabular}{lccc} 
\hline
Component   & $r$ [arcsec] & $v_\mathrm{exp}$ [km~s$^{-1}$]  & $\tau_\mathrm{k}$ [yr] \\
\hline
main cavity & 13.5$\times$18.4 & 25   & 6,500 $\pm$ 900 \\
NW lobe     & 8                & 14.6 & 4,900 $\pm$ 600 \\
NE lobe     & 7                & 15.5 & 4,000 $\pm$ 500 \\
SW lobe     & 11.3             & 27.7 & 3,700 $\pm$ 500 \\
SE lobe     & 9.8              & 26   & 3,800 $\pm$ 600 \\
\hline
\end{tabular}
\label{tab:ages}
\end{table}

\section{Discussion}

The high-quality NOT images of NGC\,40 presented here allowed us to unveil the ionization structure of this PN with unprecedented detail. 
There are noticeable similarities between the H$\alpha$, H$\beta$, [N\,{\sc ii}], [S\,{\sc ii}] and [O\,{\sc ii}] images, but clear differences when compared with the [O\,{\sc i}] and [O\,{\sc iii}] narrow-band images. 
We corroborated previous findings that the [O\,{\sc iii}] emission is contained within the cavity mapped by the H$\alpha$ emission line \citep[see][and references therein]{Meaburn1996}, whereas the [O\,{\sc i}] emission is the most extended. This stratification can be easily explained by a ionization structure, with the highest ionized material (O$^{++}$) inside. 

Moreover, NGC\,40 appears to be composed by clumps and filaments and this turbulent pattern seems to have different origins depending on the ionization characteristics of the material. The clumps and filaments in the [O\,{\sc i}], [O\,{\sc ii}] and [O\,{\sc iii}] images are not similar between them and have different shapes and sizes. To assess these morphological differences, we used the {\sc astrodendo} python package\footnote{For details see \url{https://dendrograms.readthedocs.io/en/stable/}} which provides a way to quantify the clumps and filaments in 2D images. The analysis of the three oxygen images resulted in averaged clump sizes of 13.9, 24.0 and 16.9~arcsec$^{2}$ for the [O\,{\sc i}], [O\,{\sc ii}] and [O\,{\sc iii}] images, respectively. Such morphological differences seem to be a consequence of the wind-wind interaction model of formation of PNe. As stated before, the fast wind from HD\,826 ($\sim$1000~km~s$^{-1}$) is slamming the slow wind ejected in a previous evolutionary phase producing hydrodynamical instabilities that can be seen as clumps in the O$^{++}$ image. These clumps and filaments prevent the material in the outer shells to be uniformly photoionized, creating streaks of alternate ionization regions that are detected in optical images \citep[see, Fig.~1 in][]{Toala2019}, causing the outer layers to be partially ionized and turbulent.

The $n_\mathrm{e}$ map obtained using the narrow-band [S\,{\sc ii}] images suggests a gradient from the SW region of NGC\,40 towards the SE region. We have inspected the independent [S\,{\sc ii}] narrow-band images to look for possible inconsistencies that could produce this gradient, but we did not find any notorious effect. 
On the other hand, the $n_\mathrm{e}$ map presented by \citet{LealF2011} suggest maximum values associated with the western and southern regions of the main cavity, but we note that their map has a lower spatial resolution than ours as a result of their sparse spatial coverage and interpolation technique.
Otherwise these authors report values of $n_\mathrm{e}$= 1000--3000 cm$^{-3}$, which are consistent with our results. 
The inner cavity of NGC\,40 seems to have averaged $n_\mathrm{e}=2000$~cm$^{-3}$, with the highest values of 3500~cm$^{-3}$ at the SW regions.

\subsection{Multiple ejections in NGC\,40}

The optical data presented here suggest that HD\,826, the CSPN of NGC\,40, has experienced a number of successive mass loss ejections. The NOT images of Fig.~\ref{fig:figt}, along with the resultant ratio images presented in Fig.~\ref{fig:ratios}, unveil the presence of blowout structures protruding from the main cavity of NGC\,40. These seem to correspond to two pairs of bipolar ejections aligned with PA of 80$^{\circ}$ and 135$^{\circ}$ corresponding to Axis 1 and 2 as defined in the bottom-right panel of Fig.~\ref{fig:figt}, respectively. The ejections along Axis~1 seem to have pierced through NGC\,40, but the ejections aligned along Axis~2 appear to be currently disrupting the main cavity producing small blisters structures. Surprisingly, the diffuse X-ray emission filling NGC\,40 also have two maxima along these directions.

In accordance with \citet{Meaburn1996}, we neither detect any hint of velocity pattern from the jet-like features 1N, 1S, 2N nor 2S (see Fig.~\ref{fig:figt}).
The increase by a factor of $\sim$3 of the [O\,{\sc iii}]/H$\alpha$ ratio along the direction of the southern jet-like feature disclosed by the map presented in Fig.~\ref{fig:ratios} (top-left panel) might be attributed to the effects of a jet piercing the SW cap of NGC\,40 related to feature 1S. 
If the 1S structure were indeed a jet, it had to be contained within the plane of the sky.

Its possible counterpart, the 1N feature, is even more difficult to evaluate. We also do not find any velocity pattern for the structures defined as 2N--2S in Fig.~\ref{fig:figt}. 
In fact, these features are reproduced by our {\sc shape} model as the intersection of the two pairs of northern and southern caps from the main barrel-like cavity of NGC\,40.

Finally, there are two high velocity features detected in the [O\,{\sc iii}] WHT UES spectrum located inside the main cavity of NGC\,40. These are detected at velocities $\approx$60~km~s$^{-1}$ with no apparent counterparts in blue velocities. These inner high-velocity structures resemble the clumps detected in the IR images of NGC\,40 presented by \citet{Toala2019}.
It is quite possible that these structures are actually part of a quadrupolar configuration (two pairs of jets) but their blue counterparts would be self-absorbed by the nebula. 
Alternatively we do not detect these high velocity components due to the limited depth of the observations. 
Similar velocities have been detected from internal hydrogen-poor clumps in born-again NPs A\,30, A\,78 and HuBi\,1 \citep{Fang2014,Meaburn1998,Rechy2020}. It is accepted that these clumps have been ejected during the VLTP, which reinforces the suggestion of \citet{Toala2019} that NGC\,40 belongs to this rare class of PNe.

\subsection{Formation scenario of NGC\,40}

The spatio-kinematic model presented in Section~\ref{sec:shape} can be used to envisage the formation history of NGC\,40. Since we have the expansion velocity $v_\mathrm{exp}$ of the different structures in NGC\,40 we can calculate their kinematic ages by following the equation:
\begin{equation*}
    \tau_\mathrm{k} \approx 4744 \left( \dfrac{r}{\mathrm{arcsec}} \right) \left(\dfrac{d}{\mathrm{kpc}}\right)\left(\dfrac{v_\mathrm{p}}{\mathrm{km~s^{-1}}}\right)^{-1}~\mathrm{yr},
\end{equation*}
\noindent where $r$ is the radius of each structure, $d$ is the distance to NGC\,40 and $v_\mathrm{p}$ is the deprojected rate of expansion. Since the inclination of NGC\,40 with respect to the plane of the sky is virtually zero,  $v_\mathrm{p} \approx v_\mathrm{exp}$. The estimated distance using data from Gaia data to NGC\,40 is 1.87~kpc \citep{BJ2018}. The estimated kinematic ages of each component are listed in Table~\ref{tab:ages}.

The barrel-like main cavity of NGC\,40 is the oldest structure in this PN with an kinematical age of almost 6500~yr. 
Other determinations of kinematic ages in the literature are consistent with this value. 
\citet{Koesterke1998} report a kinematic age for the nebula of $\sim$4,000~yr considering a distance of 986 pc and \citet{Monteiro2011} find an age of 3500$\pm$500~yr using a distance of 1.28 kpc.

The four caps in our morpho-kinematic model of NGC\,40 have slightly different kinematical ages with the northern caps having larger $\tau_\mathrm{k}$ values. However, considering the errors in their age determinations, these were probably formed by two contemporaneous bipolar ejections $\approx$4100$\pm$550~yr ago. 
This situation is common in binary scenarios \citep[see][]{GuerreroRings2020}. Numerical models that simulate bipolar PN formation using common envelope theory confirm it.
For example, \citet{Zou2020} reported that, when considering cooling to the low-momentum flow in an adiabatically evolving lobe, we will obtain asymmetric bipolar nebulae where one lobe will have a greater extension, as observed in NGC\,40.

The rest of the bipolar ejections discovered and reported in this paper are younger than the kinematic ages of the main cavity and lobes in NGC\,40. In particular, the clumps detected in the WHT UES [O\,{\sc iii}] spectrum were ejected approximately 2000--2600~yr ago, which makes their formation by VLTP feasible and adding NGC\,40 to the selected group of born-again PNe.

Considering all the structures unveiled by the present work, we can undoubtedly say that NGC\,40 has experienced several episodes of mass ejection, making it a more complicated object than previously thought. 
Ring-like structures in the halo of NGC\,40 have been found through optical and IR observations \citep{Corradi2004,RamosLarios2011}, which can be interpreted as evidence of a binary system at the core of NGC\,40 \citep[see, e.g.,][]{Mastrodemos1999}. 
Binarity is the most likely formation mechanism to explain most features in PNe \citep{Soker1997,DeMarco2009}, even scenarios in which the binary system enters a common envelope evolution \citep[e.g.,][]{Ivanova2013,GS2021,Zou2020}. However, the multiple ejections with (at least) 4 different axis challenge these ideas. Other possibility could be the presence of a triple system \citep[e.g.,][and references therein]{Soker2021}, but it is expected that such systems produce more messy PNe, which is not the case of NGC\,40.

There are other clear examples of PNe that still challenge our understanding of PNe formation mechanisms. 
For example, IC\,4593 harbours several jet-like clumps along orthogonal directions \citep{Corradi1997}. 
Recently, \citet{Henney2021} thoroughly demonstrated that NGC\,6210 has 5 symmetry axes and clear asymmetries. Objects like these, including now NGC\,40, urges us to further peer into the formation mechanisms of PNe 
and very specifically on the chaotic final phases of the binary evolution through a common envelope.

\section{SUMMARY AND CONCLUSIONS}

We have presented the analysis of narrow-band images and high-resolution spectra of the Bow-tie Nebula, a.k.a. NGC\,40, to produce a morpho-kinematic study of this PNe. We found a complex distribution of the ionization structure produced by turbulent structures caused by a combination of hydrodynamical and shadowing instabilities.

Our morpho-kinematic model of NGC\,40 obtained with {\sc shape} helped us to assess the kinematic ages of the major morphological components of this PNe. At an adopted distance of 1.9~kpc, the main cavity has a kinematic age of 6500~yr. The northern and southern lobes, which are modelled with cap-like structures, seem to have been ejected 4200~yr ago.

We found four jet-like ejections aligned in different directions. Our images and spectra showed that some of these jet-like features have pierced NGC\,40 in four regions at the main cavity, with an extra jet-like feature that has pierced the SE lobe. The youngest ejections are detected inside the main cavity of NGC\,40 with higher velocities of 60~km~s$^{-1}$. The youngest internal ejections seem to be in line with the VLTP event experienced recently by this PN proposed by \citet{Toala2019}.

We conclude that NGC\,40 had a complex formation scenario involving multiple mass-loss episodes, some of them even not aligned with the main symmetry axis of this nebula. 
Multi-axis PNe such as IC\,4593, NGC\,40 and NGC\,6210 challenge the current paradigm of PN formation.

\section*{Acknowledgements}

The authors thank the referee for comments and suggestions that improved the presentation of our results.
J.B.R.G. thanks Consejo Nacional de Ciencias y Tecnolog\'{i}a (CONACyT, Mexico) for
research student grant. J.B.R.G. and J.A.T. thank funding by Direcci\'{o}n General de Asuntos
del Personal Acad\'{e}mico (DGAPA) of the Universidad Nacional
Aut\'{o}noma de M\'{e}xico (UNAM) project IA100720. J.A.T. thanks the Marcos Moshinsky Foundation (Mexico). J.A.T. and G.R.-L. acknowledge support from CONACyT (grant 263373). LS acknowledges a grant UNAM-PAPIIT IN101819 and IN110122 (Mexico). M.A.G. is funded by the Spanish Ministerio de Ciencia, Innovaci\'on y Universidades (MCIU) grant PGC2018-102184-B-I00, co-funded by FEDER funds. M.A.G. acknowledges support from the State Agency for Research of the Spanish MCIU through the ‘Center of Excellence Severo Ochoa’ award to the Instituto de Astrof\'\i sica de Andaluc\'\i a (SEV-2017-0709). J.A.L acknowledges a grant DGAPA-PAPIIT IN103519 (Mexico) CONACyT A1-S-15140. 

The data presented here were obtained in part with ALFOSC, which is provided by the Instituto de Astrofisica de Andalucia (IAA) under a joint agreement with the University of Copenhagen and NOT. Partially based upon observations carried out at the Observatorio Astron\'{o}mico Nacional on the Sierra San Pedro M\'{a}rtir (OAN-SPM), Baja California, Mexico. The authors thank the telescope operator P.F.\,Guill\'{e}n for valuable guidance during several observing runs.
This work has made extensive use of NASA's Astrophysics Data System.

\section*{DATA AVAILABILITY}

The data underlying this article will be shared on reasonable request
to the corresponding author.





\end{document}